\documentclass[english,letterpaper,prb,twocolumn,altaffilletter,nofootinbib]{revtex4}
\usepackage[T1]{fontenc}
\usepackage[latin9]{inputenc}
\usepackage{babel}
\usepackage{graphicx}
\usepackage[unicode=true]
 {hyperref}

\makeatletter

\providecommand{\tabularnewline}{\\}

\@ifundefined{textcolor}{}
{%
 \definecolor{BLACK}{gray}{0}
 \definecolor{WHITE}{gray}{1}
 \definecolor{RED}{rgb}{1,0,0}
 \definecolor{GREEN}{rgb}{0,1,0}
 \definecolor{BLUE}{rgb}{0,0,1}
 \definecolor{CYAN}{cmyk}{1,0,0,0}
 \definecolor{MAGENTA}{cmyk}{0,1,0,0}
 \definecolor{YELLOW}{cmyk}{0,0,1,0}
}

\makeatother

\begin{document}

\title{Simulating Interference and Diffraction in Instructional Laboratories}

\author{Leon N. Maurer}

\email{lnmaurer@wisc.edu or leon.maurer@gmail.com}

\selectlanguage{english}%

\affiliation{Department of Physics, University of Wisconsin-Madison, Madison,
Wisconsin 53705}

\date{09/20/2012}
\begin{abstract}
Studies have shown that standard lectures and instructional laboratory
experiments are not effective at teaching interference and diffraction.
In response, the author created an interactive computer program that
simulates interference and diffraction effects using the Finite Difference
Time Domain method. The software allows students to easily control,
visualize, and quantitatively measure the effects. Students collected
data from simulations as part of their laboratory exercise, and they
performed well on a subsequent quiz---showing promise for this approach.
\end{abstract}
\maketitle

\section{Introduction}

Often, interference and diffraction is taught by showing the mathematics
in lecture and then performing laser interference and diffraction
experiments in an instructional laboratory. However, studies show
this teaching method is lacking.

\begin{figure}
\begin{centering}
\includegraphics[width=1\columnwidth]{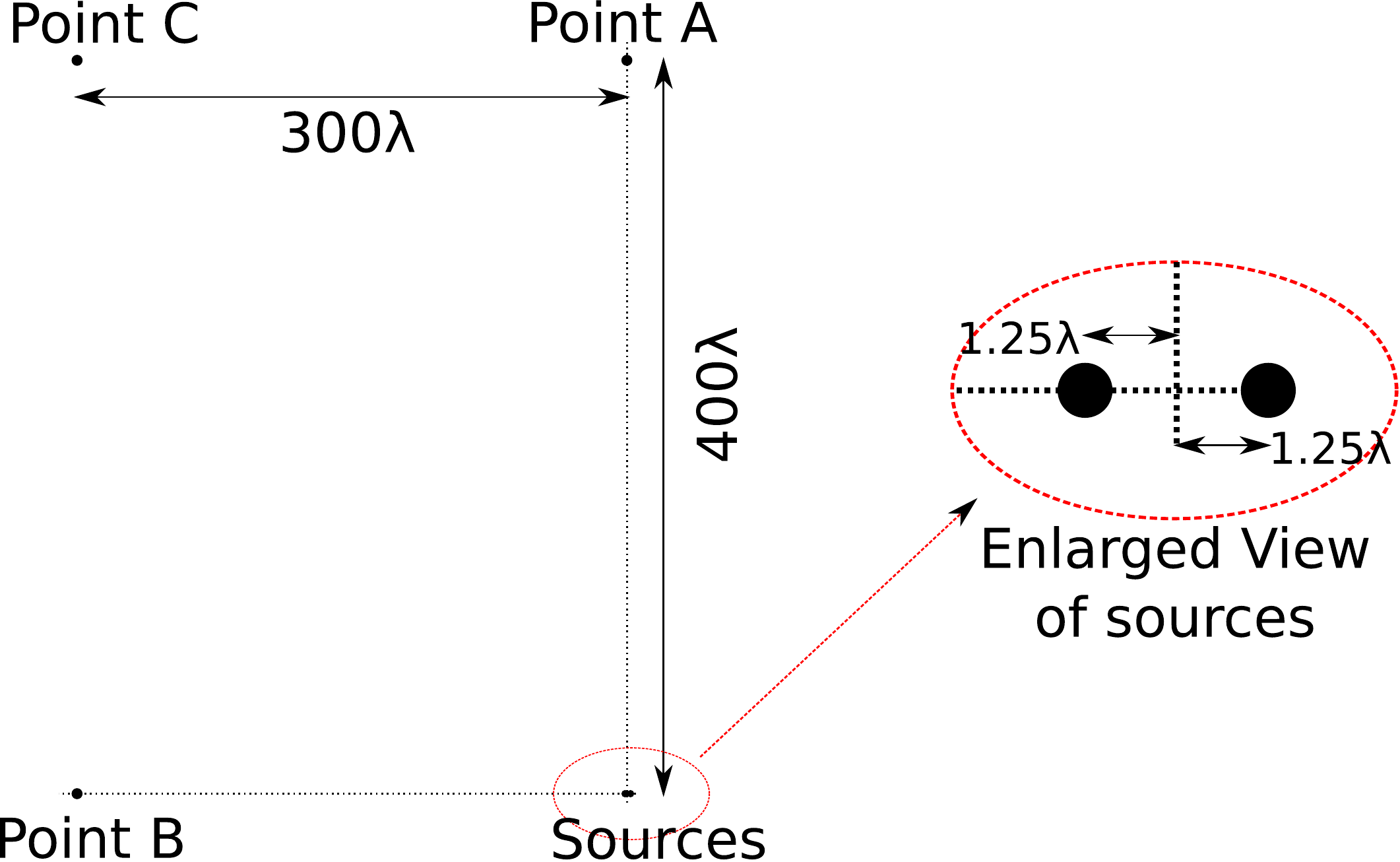}
\par\end{centering}

\caption{The diagram for the interference quiz.\label{fig:quiz}}
\end{figure}

Consider the following problem (see Figure \ref{fig:quiz}), which
should be straightforward for a student who understands the concepts.
Two point sources, $2.5$ wavelengths ($\lambda$) apart, generate
waves in phase. Is there constructive interference, destructive interference,
or neither at points A, B, and C?

A University of Washington study asked this and other questions to
$\approx1200$ undergraduates in their standard, calculus-based, introductory
physics course. Only $\approx35\%$ answered correctly for both points
A and B, and only $\approx5\%$ answered correctly for point C. Graduate
students also performed poorly; only $\approx25\%$ answered correctly
for point C.\cite{wosilait:S5} These mistakes were primarily due
to fundamental misunderstandings; post-quiz student interviews included
responses like, \textquoteleft{}\textquoteleft{}I suppose that $2.5\lambda$
{[}the slit separation{]} is small compared to $400\lambda$ and $300\lambda$,
so the sources here act like a single source.\textquoteright{}\textquoteright{}\cite{ambrose:146}
Moreover, the typical instruction method did not help; the scores
for point C before and after lecture and laboratory were basically
unchanged.\cite{ambrose:146,wosilait:S5}

Part of the problem is that light's wave nature is only visible indirectly---though
effects like interference and diffraction. Light's peaks and valleys
cannot be viewed directly. Hands-on experience seeing and controlling
the waves' interactions is useful for student understanding, so a
better wave medium is needed. Ripple tanks provide this, but there
are significant trade-offs between easy of use, cost, and measurement
accuracy.

For example, Wosilait et al. provided students with simple ripple
tanks---pans, dowels, and sponges. They are cheap and easy to use
but do not allow quantitative measurements. Still, they proved effective
as part of an intensive tutorial system that raised scores for point
C to $\approx45\%$.\cite{wosilait:S5}

Measurements can be performed with a ripple tank---at the price of
increased cost and complexity. They require a strobe device and a
consistent wave source---not a student rolling a dowel back and forth.
The strobe can either be a light\cite{pasco} (easier but more expensive)
or a spinning, slotted disk\cite{strobe} (cheaper but more complicated).
Either makes the waves appear frozen, making their measurement easier.

Computer simulations can also display waves, and they have potential
advantages over ripple tanks. Simulations can allow finer control
and more accurate measurements. They are available for free. They
can be used outside of laboratory. And they allow better visualization
by avoiding unwanted reflections, being easily paused, and having
superior contrast.

Dozens of simulations are available (search \url{http://www.merlot.org/}).
Some display what would be seen in a laser experiment---without showing
the waves interacting.\cite{noVisEz} Others are non-interactive animations.\cite{animated}
Some have interactive displays.\cite{Physlets} A few allow measurement
of the instantaneous wave height/field strength.\cite{webTop}

Those simulations are primarily used for qualitative understanding.
However, I wanted a simulation that could answer questions like the
aforementioned one---albeit on a smaller scale---and could thereby
test interference conditions. This requires easily measured amplitudes
at any user desired grid point. No simulations combined that with
interactive, easy to control, animated waves.

Similar simulations have been proposed, but their use in instructional
laboratories has not been reported. Frances et al. proposed using
a Finite Difference Time Domain (FDTD) electromagnetics simulation
to show interference and diffraction effects, but they have only reported
its use in lecture demonstrations.\cite{FDTDGUI} Werley et al. also
suggested using FDTD simulations in laboratories, but they instead
used pre-recorded videos of actual propagating radiation.\cite{DirectVisualization}

So, I wrote an FDTD simulation that allows easy and accurate measurements,
and I constructed a laboratory exercise that uses the simulation for
quantitative measurements, not just qualitative understanding.

\section{The Program}

The FDTD technique solves differential equations by discretizing them
in both space and time. It has proved popular and effective for simulating
electromagnetics, and there are many fine references on the technique.\cite{1138693,TafloveHagness,FDTD}
So, the equations are not reproduced here;%
\footnote{For reference, the program simulates a $TM_{z}$ wave using the standard
Yee lattice and second-order-accurate finite difference approximations.%
} this section summarizes aspects of the simulation and interface relevant
to its use.

Figure \ref{fig:interface} shows the program's interface, which has
five plots. The two large plots show $E_{z}$ (upper plot) and $E_{zRMS}$
(lower plot). For both, black is the smallest value and white is the
largest value, with shades of gray in between. The three smaller plots
show $E_{z}$ and $\pm\sqrt{2}E_{zRMS}$---an envelope for $E_{z}$---along
the horizontal and vertical dashed lines through the two larger plots.
Those lines can be moved with the keyboard or mouse, and $x$, $y$,
$E_{z}$, and $E_{zRMS}$ at those lines' intersection is displayed
in the center right area between the two vertical plots. Knowing $E_{zRMS}$
at that point allows users to home in on extrema.

A plane wave---with a wavelength of $20$ grid cells---enters from
the left. It is not simulated but calculated analytically. At the
start of the simulation, the wave's magnitude is ramped up gradually
to avoid potentially unstable high frequency components.

The barrier---the red line visible in both large plots---is a perfect
conductor, and the openings in the barrier are hard sources that inject
the incoming wave in to the FDTD domain---the area to the right of
the barrier. Openings in the barrier can be added, removed, and modified
using the barrier control frame---at the right of the interface.

Split-field perfectly matched layers\cite{Berenger1994185,TafloveHagness}
terminate the other three sides of the FDTD domain. These boundaries
reduce reflections to imperceptible levels, effectively giving the
simulation open boundaries.

When the simulation starts or the barrier is modified, a timer appears
over the $E_{zRMS}$ plot---counting down until steady state is reached.
Afterwards, $E_{zRMS}$ is reset to remove transients, and another
countdown appears for one wave time period. The steady state $E_{zRMS}$
is calculated by averaging over that time.

Among its other features, the simulation also has a fast forward mode,
which saves time by not updating the plots. In that mode, the simulation
runs until the current countdown is done.

The laboratory's computers---running Windows 7 with Intel Core 2 Duo
processors---take $\approx55ms$ per timestep, resulting in a smooth
animation.

The software is written in Python using NumPy for the calculations,
TkInter for the interface, and the Python Imaging Library for the
plots. Those libraries are available for Windows, OS X, and Linux.
Executables, source code, and program information are available at
\url{http://lnmaurer.github.com/Interference-Inference-Interface/};
the program's source code available under the GNU Public License version
2.

\section{The Simulations}

The laboratory was tailored for a particular class---aimed at future
physics majors, but tweaking the exercises for other courses should
be straightforward. Besides simulations, the laboratory also included
pen and paper work and short laser/slit experiments.%
\footnote{The worksheet is available at \href{https://github.com/lnmaurer/Interference-Diffraction-Worksheet}{https://github.com/lnmaurer/Interference-Diffraction-Worksheet}.%
} However, four simulations are at the laboratory's heart: of narrow
single, double, and triple slits and a wide single slit.

\subsection{Narrow Single Slit}

\begin{figure}
\begin{centering}
\includegraphics[width=1\columnwidth]{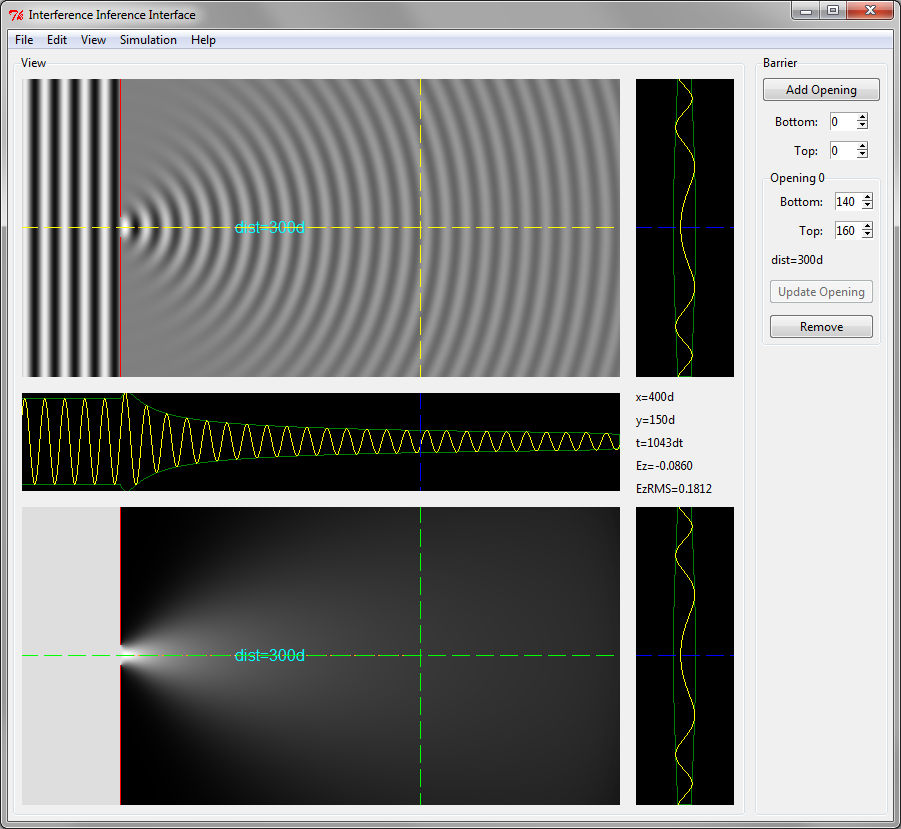}
\par\end{centering}

\caption{The single narrow slit simulation.\label{fig:single}\label{fig:interface}}
\end{figure}

See Figure \ref{fig:single}. This simulation provides a baseline
for comparison. It shows that a single narrow opening is not responsible
for the interference patterns seen in the following simulations.

The course's lectures included a mathematical description of point
sources in 3D, so this simulation introduced students to the asymmetric
sources used in the laboratory. To acquaint students with the simulation's
controls, I had them roughly measure how fast the amplitude decreased
with distance from the opening; because this is a 2D simulation, the
amplitude falls off less quickly than in 3D.

\subsection{Narrow Double Slits}

\begin{figure}
\begin{centering}
\includegraphics[width=1\columnwidth]{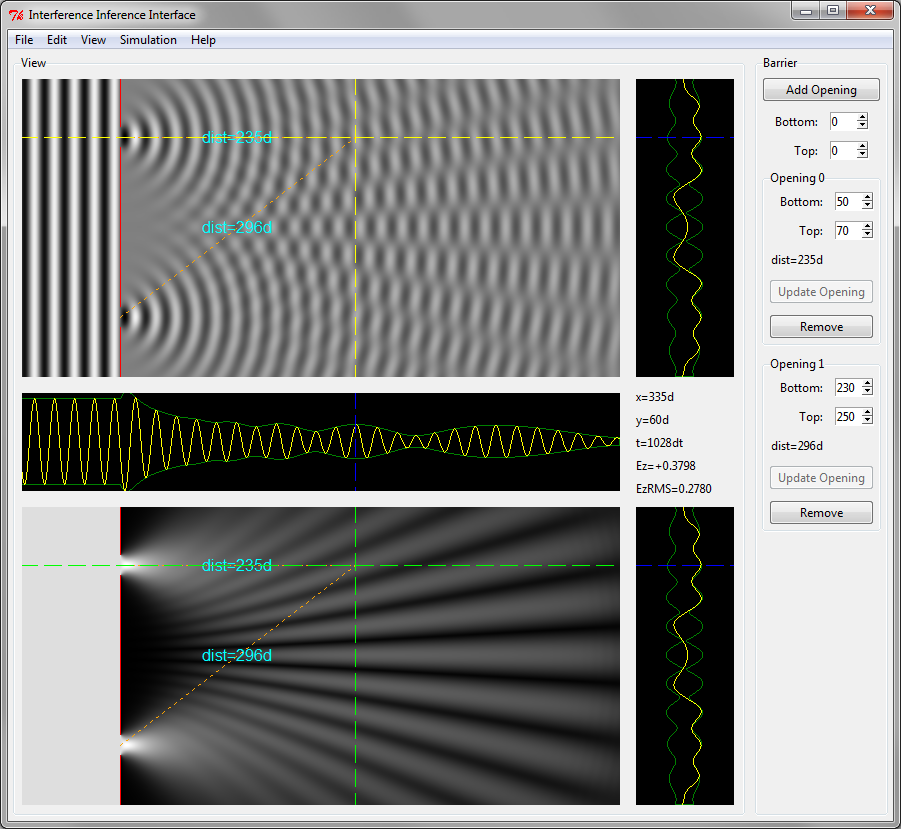}
\par\end{centering}

\caption{The double narrow slit simulation with the dashed lines intersecting
at an $E_{zRMS}$ maximum.\label{fig:double}}
\end{figure}

See Figure \ref{fig:double}. This is the key simulation; it lets
students discover the conditions for constructive and destructive
interference. Students were asked to find four maxima and four minima
of $E_{zRMS}$---in either $\hat{x}$ or $\hat{y}$ and in the right
half of the domain, calculate $\frac{\left|d_{0}-d_{1}\right|}{\lambda}$
for each---where $d_{0}$ and $d_{1}$ are the distances from the
extrema to the openings, and then find the pattern in those numbers.

In principle, the students already knew the conditions, but they seemed
new to many. Furthermore, most students seemed to be comprehending
the conditions for the first time.%
\footnote{Discretization in the program can slightly alter the conditions. However,
$\left|d_{0}-d_{1}\right|$ was always within one. For example, the
point where the dashed lines crosses is a maximum in Figure \ref{fig:double},
and $\frac{\left|d_{0}-d_{1}\right|}{\lambda}=\frac{\left|235-296\right|}{\lambda}=\frac{61}{20}=3.05$.
These small errors were not problematic for students.%
} Pen and paper work followed to reinforce their understanding. This
included drawing a couple waves that resulted in those conditions
(using a simplified version of\cite{drawing1,drawing2}) and mathematically
verifying that the conditions lead to waves in or out of phase.

\subsection{Narrow Triple Slits}

\begin{figure}
\begin{centering}
\includegraphics[width=1\columnwidth]{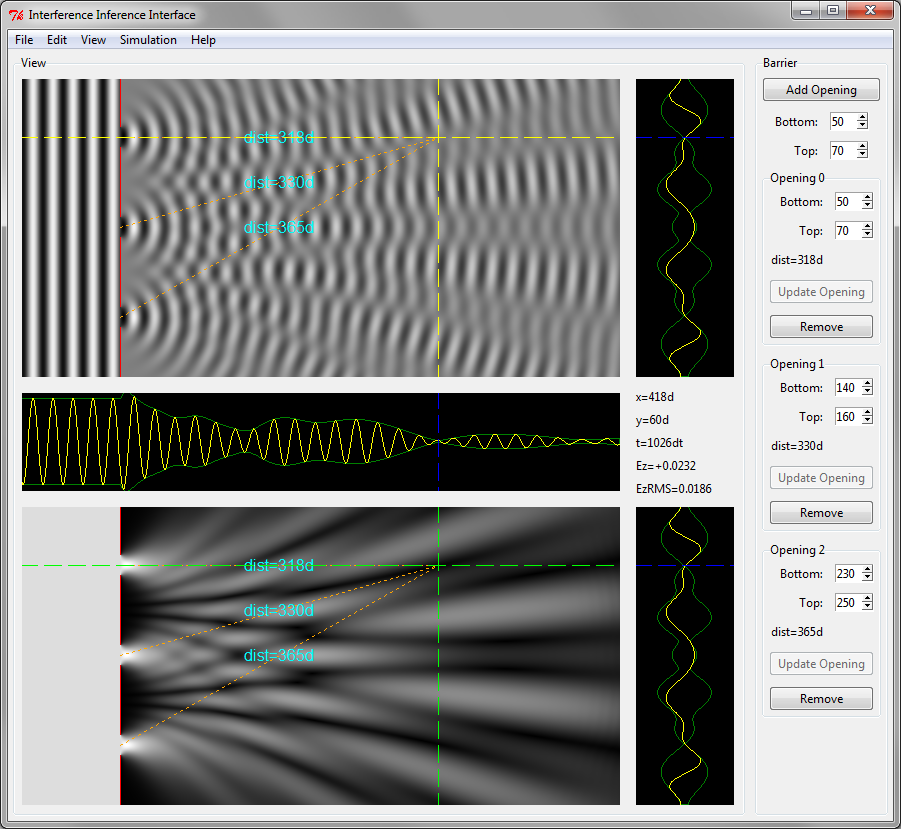}
\par\end{centering}

\caption{The triple narrow slit simulation with the dashed lines intersecting
at an $E_{zRMS}$ minimum. It is located $15.9\lambda$, $16.5\lambda$,
and $18.25\lambda$ from the three openings.\label{fig:triple}}
\end{figure}

See Figure \ref{fig:triple}. The additional slit makes students closely
consider the logic behind the extrema conditions.

Here, the students were asked to find points that were simultaneously
extrema in both $\hat{x}$ and $\hat{y}$. This is straightforward
with the simulation, but---since this requires fine control in both
$\hat{x}$ and $\hat{y}$---this would be difficult using the standard
experimental setup.

The condition for constructive interference still holds. To get a
maximum, all pairs should be maximized.

However, the minimization condition is more complicated.\cite{Ovchinnikov200333}
The two slit condition caused the two waves to roughly cancel. That
would leave the third wave undiminished. Additionally, it is mathematically
impossible for all differences in the distances to be a half integer
number of wavelengths.

This is an instructive example that is missing from laboratories that
do not use simulations for quantitative measurements.

\subsection{Wide Single Slit}

\begin{figure}
\begin{centering}
\includegraphics[width=1\columnwidth]{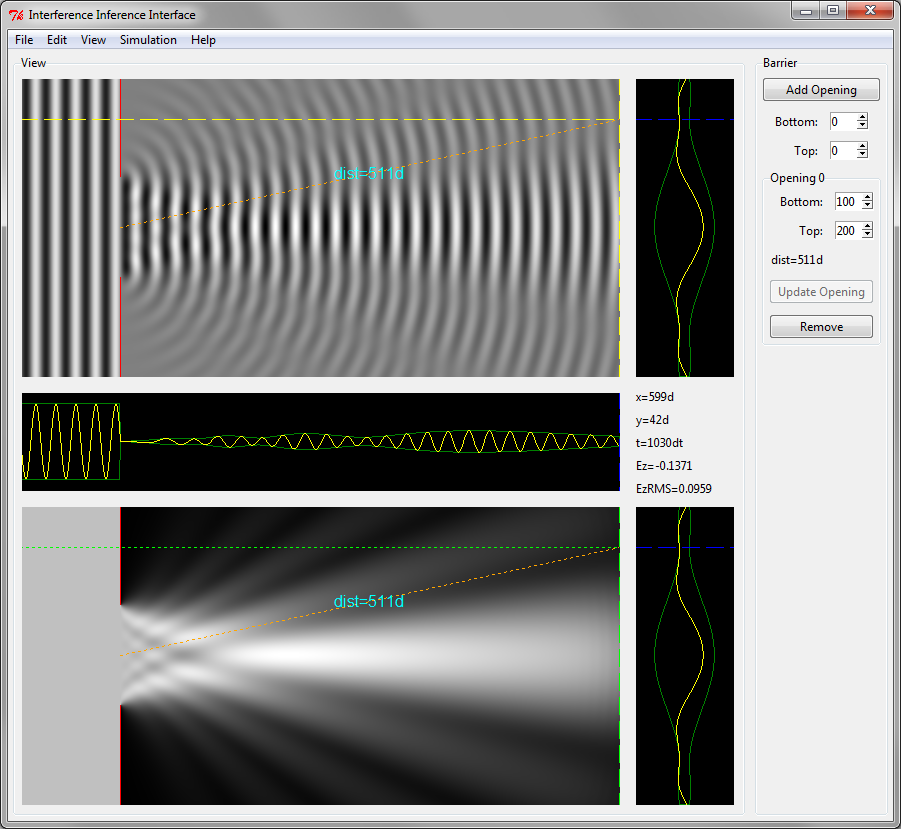}
\par\end{centering}

\caption{The wide single slit simulation with the dashed lines intersecting
at the first minimum. It occurs at $\theta=\arctan\left(\frac{150-42}{599-100}\right)=0.213$,
whereas the far field limit has $\theta=\frac{20}{100}=0.2$.\label{fig:wide}}
\end{figure}

See Figure \ref{fig:wide}. The right side of the domain gets close
to the far field limit. While the domain is too small to truly display
far field effects---a limitation common to many ripple tanks, the
simulation successfully shows diffraction effects arising from waves.
The simulation could be used to investigate near field effects---something
difficult with normal instructional laboratory equipment.

\section{Results}

\begin{table}[htbp]
\caption{Scores for point C. The first three columns of results are from Wosilait
et al. and are rounded to the nearest 5\%.\cite{wosilait:S5}\label{tab:results}}

\centering{}%
\begin{tabular}{|c|c|c|c|c|}
\hline 
 & Old UWash & UWash Grad & New UWash & My 15\tabularnewline
\hline 
\hline 
Right Method & 10\% & 55\% & 70\% & 73\%\tabularnewline
\hline 
and Right Answer & 5\% & 25\% & 45\% & 60\%\tabularnewline
\hline 
\end{tabular}
\end{table}

Only $15$ students used the finial version of the laboratory, so
the only conclusion I can draw with confidence is that students enjoyed
it; I collected anonymous feedback, and it was overwhelmingly positive.
Still, for a rough gauge of the method's effectiveness, I quizzed
the students with questions from the University of Washington study.
Scores for point C are in Table \ref{tab:results}.

\begin{figure}
\begin{centering}
\includegraphics[width=1\columnwidth]{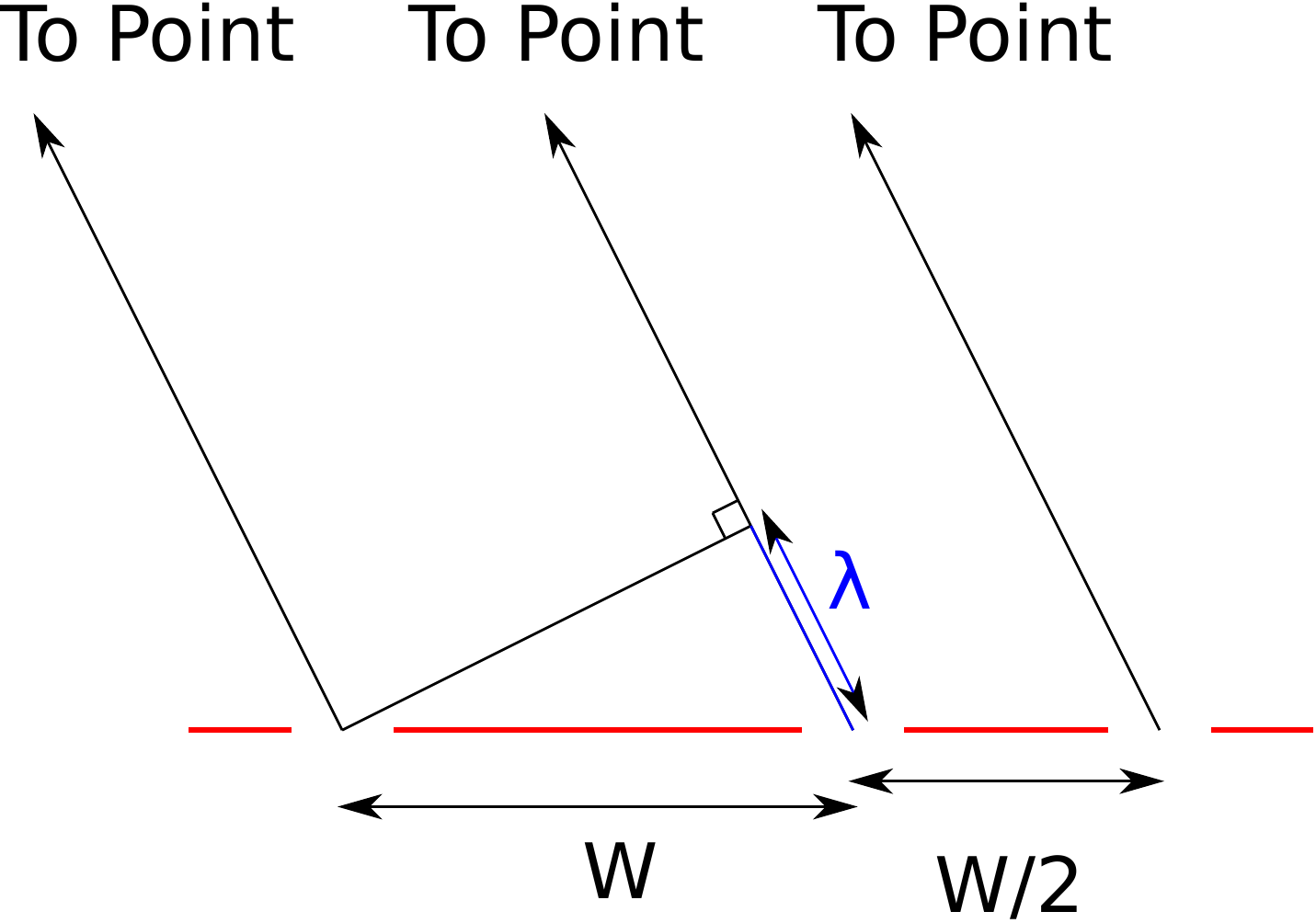}
\par\end{centering}

\caption{Figure for the second quiz question.\label{fig:quiz2}}
\end{figure}

They did well on all parts of the quiz, relative to others. $60\%$
of the students answered correctly for point C. $80\%$ correctly
identified the behavior at both points A and B---students in the old
University of Washington system scored $35\%$.\cite{wosilait:S5}
The quiz included another question from the University of Washington
study. See Figure \ref{fig:quiz2}. What would be observed at the
far away point---first with just the left two slits and then with
all three slits? $80\%$ got both parts right.

\section{Conclusion}

Interference and diffraction are tricky subjects to teach, but this
approach shows promise in making them more accessible to students.
The simulation allows them to easily control, visualize, and quantitatively
measure interference and diffraction effects. That makes the software
a powerful tool.

The progression of simulations described here appears effective. First,
a single narrow slit familiarized students with the simulation and
showed more was needed for interference. Then, a double slit simulation
showed Young's experiment in a new light, with visible waves and easy
measurements. Next, a triple slit experiment made students reconsider
the constructive and destructive interference conditions; one held
but the other fell. Finally, a diffraction simulation showed that
similar effects could arise from one wide slit.

While the sample size here was small, the results are encouraging.
I hope eventually to try this approach with more students in more
classes.

It would be difficult to experimentally mimic some of the simulations
with instructional laboratory equipment. The simulation could also
easily demonstrate more advanced topics with minimal modification.
The permittivity, permeability, and conductivity are definable at
every point in the simulation domain. Changing those parameters would
allow the simulation to demonstrate waveguides, reflection due to
changes in dielectric constants, etc. Simulations of a modified two
slit experiment---where the upper and lower halves of the domain have
different indices of refraction---could illustrate more conceptual
topics, like the difference between physical and optical path length.\cite{opticalPath}
\begin{acknowledgments}
I would like to thank Dr. James Reardon for encouragement during this
laboratory's development, Professor Susan Hagness for feedback on
the simulation, and Professors Daniel Chung and Lisa Everett for letting
me try this approach in their course, where I was the teaching assistant.
\end{acknowledgments}
\bibliographystyle{unsrt}
\bibliography{cleanbib}

\end{document}